Reduced steroid activation of elephant shark glucocorticoid and mineralocorticoid receptors after inserting four amino acids from the DNA-binding domain of lamprey corticoid receptor-1

Yoshinao Katsu[1,2,] *, Jiawen Zhang[1], Michael E. Baker[3,4,] *

[1] Faculty of Science

Hokkaido University

Sapporo, Japan

[2] Graduate School of Life Science

Hokkaido University

Sapporo, Japan

[3] Division of Nephrology-Hypertension

Department of Medicine, 0693

University of California, San Diego

9500 Gilman Drive

La Jolla, CA 92093-0693

Center for Academic Research and Training in Anthropogeny (CARTA) [4]

University of California, San Diego

La Jolla, CA 92093

*Correspondence to

Y. Katsu; E-mail: ykatsu@sci.hokudai.ac.jp

M. E. Baker; E-mail: mbaker@health.ucsd.edu

**Abstract** Atlantic sea lamprey contains two corticoid receptors (CRs), CR1 and CR2, that are identical except for a four amino acid insert (Thr-Arg-Gln-Gly) in the CR1 DNA-binding domain (DBD). Steroids are stronger transcriptional activators of CR2 than of CR1 suggesting that the insert reduces the transcriptional response of lamprey CR1 to steroids. The DBD in elephant shark mineralocorticoid receptor (MR) and glucocorticoid receptor (GR), which are descended from a CR, lack these four amino acids, suggesting that a CR2 is their common ancestor. To determine if, similar to lamprey CR1, the presence of this insert in elephant shark MR and GR decreases transcriptional activation by corticosteroids, we inserted these four CR1-specific residues into the DBD of elephant shark MR and GR. Compared to steroid activation of wild-type elephant shark MR and GR, cortisol, corticosterone, aldosterone, 11-deoxycorticosterone and 11-deoxycortisol had lower transcriptional activation of these mutant MR and GR receptors, indicating that the absence of this four-residue segment in the DBD in wild-type elephant shark MR and GR increases transcriptional activation by corticosteroids.

**Introduction**

The sea lamprey (*Petromyzon marinus*) belongs to an ancient group of jawless vertebrates known as cyclostomes, which are basal vertebrates that evolved about 550 million years ago (1–4). Sea lamprey contains a corticoid receptor (CR) (5–8), which belongs to the nuclear receptor family of transcription factors (9,10), which also contains the mineralocorticoid receptor, glucocorticoid receptor, progesterone receptor, estrogen receptor and androgen receptor (9–13). A CR in an ancestral cyclostome is the common ancestor to the mineralocorticoid receptor (MR) and the glucocorticoid receptor (GR) in vertebrates (5,7). The MR and GR first appear as separate steroid receptors in cartilaginous fish (5–7,14,15). These two closely related steroid receptors regulate important physiological responses in vertebrates. The MR regulates electrolyte transport in the kidney and colon (12,16–21), as well as regulating gene transcription in a variety of non-epithelial tissues (22–26). The GR has diverse physiological actions in vertebrates including in development, metabolism, the stress response, inflammation and glucose homeostasis (27–31).

The importance of the MR and GR in vertebrate physiology stimulated our interest in understanding the evolution of the MR and GR from the CR (7,8,32). Unfortunately, as a result of complexities in the sequencing and assembly of the lamprey's highly repetitive and GC rich genome (3,33,34) until recently, only a partial CR sequence was available in GenBank (5). Fortunately, the recent sequencing of the lamprey germline genome (35) provided contiguous DNA encoding the complete sequences two CR isoforms, CR1 and CR2, which differ only in a novel four amino acid insertion, Thr, Arg, Gln, Gly, (TRQG) in the DNA-binding domain (DBD) (8) (Figure 1).

With the two CR sequences in hand, we characterized the response of HEK293 cells transfected with each CR isoform to corticosteroids (8) and found that the loss of these four amino acids in the DBD of CR2 increased transcriptional activation by about two-fold in the presence of corticosteroids such as 11-deoxycorticosterone, 11-deoxycortisol, cortisol, corticosterone and aldosterone.

```
lamprey CR1:       CLICSDEASGCHYGVLTCGSCKVFFKRAVEGTRQGQHNYLCAGRNDCIIDKIRRKNCPACRLRKCIQAGM
lamprey CR2:       CLICSDEASGCHYGVLTCGSCKVFFKRAVEG----QHNYLCAGRNDCIIDKIRRKNCPACRLRKCIQAGM
elephant shark GR: CLVCSDEASGCHYGVLTCGSCKVFFKRAVEG----QHNYLCAGRNDCIIDKIRRKNCPACRFRKCLQAGM
elephant shark MR: CLVCSDEASGCHYGVLTCGSCKVFFKRAVEG----QQNYLCAGRNDCIIDKIRRKNCPACRLRKCLKAGM
human GR:          CLVCSDEASGCHYGVLTCGSCKVFFKRAVEG----QHNYLCAGRNDCIIDKIRRKNCPACRYRKCLQAGM
human MR:          CLVCGDEASGCHYGVVTCGSCKVFFKRAVEG----QHNYLCAGRNDCIIDKIRRKNCPACRLQKCLQAGM
```

**Figure 1. DNA-binding domains of lamprey CR1 and CR2 and elephant shark and human MR and GR.** The DNA-binding domain of lamprey CR1 has an insertion of four amino acids that is absent in CR2 and in elephant shark and human MR and GR. Differences between the sequence of lamprey CR and elephant shark and human MR and GR are shown in red.

Elephant shark mineralocorticoid receptor (MR) and glucocorticoid receptor (GR), which are descended from the CR, lack these four amino acids in their DBD, suggesting that they evolved from a CR2 ancestor. To determine if, similar to lamprey CR2, the absence of this insert in the DBD of elephant shark MR and GR has a functional consequence, we inserted these four residues into their DBDs to convert them to a CR1-like sequence. Here we report that cortisol, corticosterone, aldosterone, 11-deoxycorticosterone and 11-deoxycortisol have lower transcriptional activation of these mutant elephant shark MR and GR receptors, indicating that compared to mutant elephant shark MR and GR, the absence of this four-residue segment in the DBD in both wild-type elephant shark MR and GR increases activation by corticosteroids.

**Results**

**Corticosteroid-dependent activation of mutated elephant shark MR and GR.**

To gain a quantitative measure of corticosteroid activation of mutated elephant shark MR and GR, we determined the concentration dependence of transcriptional activation by corticosteroids of wild-type and mutated elephant shark GR transfected into HEK293 cells with either an MMTV-luciferase promoter (Figure 2A-E) (9,36,37) or a TAT3 luciferase promoter (9,38) (Figure 2F-J). A parallel study was done with wild-type and mutated elephant shark MR (Figure 3). Luciferase levels were used to calculate an EC50 value and fold-activation for each steroid for wild-type and mutated elephant shark GR and MR (Table 1).

The results shown in Figures 2 and 3 and summarized in Table 1 reveal that addition of the TRGQ sequence to the DBD in elephant shark GR and MR reduces fold-activation of these receptors by corticosteroids in HEK293 cells transfected with either the MMTV promoter or the TAT3 promotor. Both wild-type and mutant elephant shark GR have stronger transcriptional activation than corresponding wild-type and mutant elephant shark MR, similar to that reported for wild-type human GR and MR (39) and wild-type trout GR and MR (40). For most corticosteroids, fold-activation of elephant shark GR-DBD-TRGQ was reduced by about 35% to 50% compared to wild-type elephant shark GR. The exception is cortisol activation of elephant shark GR-DBD-TRGQ in the presence of TAT3, which was reduced by 10% compared to fold-activation of wild-type elephant shark GR. At 10 nM 11-deoxycorticosterone, there was little activation of either wild-type elephant shark GR or the GR-DBD-TRGQ mutant. Corticosteroids also had lower transcriptional activation of elephant shark MR-DBD-TRGQ compared to wild-type elephant shark MR (Figure 3, Table 1).

The EC50s of corticosteroids for wild-type and mutant GR or wild-type and mutant MR were similar. The TRGQ insertion in elephant shark GR and MR mainly affected fold-activation by corticosteroids. Elephant shark GR showed greater transcriptional activation in the presence of the MMTV promoter compared to the TAT3 promoter (Figure 2, Table 1). For example, fold-activation of cortisol for elephant shark GR was 580 in the presence of MMTV and 157 in the presence of TAT3.

Elephant shark MR and elephant shark MR-DBD-TRGQ also had higher fold-activation with MMTV promoter compared to the TAT3 promoter, although it varied with the corticosteroid. For example, fold-activation of DOC for elephant shark MR was 12.4 in the presence of MMTV and 8.0 in the presence of TAT3. And fold-activation of aldosterone for elephant shark MR was 12 in the presence of MMTV and 10.7 in the presence of TAT3.

Elephant shark GR

A: MMTV-luc (Cortisol)
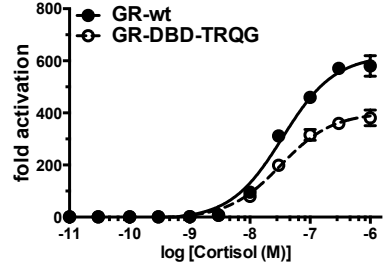

F: TAT3-luc (Cortisol)
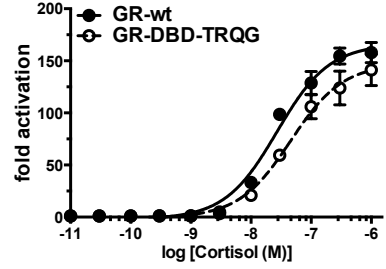

B: MMTV-luc (11-deoxycorticosterone)
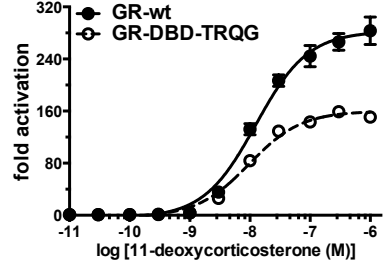

G: TAT3-luc (11-deoxycorticosterone)
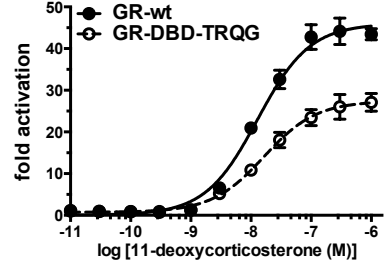

C: MMTV-luc (Corticosterone)
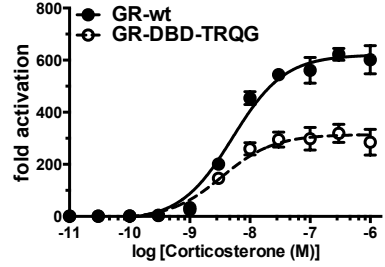

H: TAT3-luc (Corticosterone)
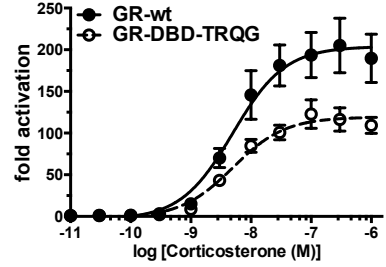

D: MMTV-luc (11-deoxycortisol)
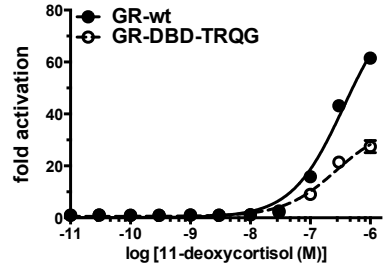

I: TAT3-luc (11-deoxycortisol)
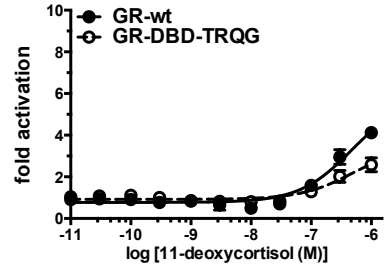

E: MMTV-luc (Aldosterone)
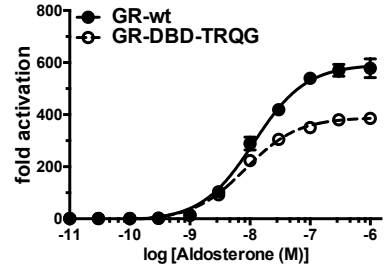

J: TAT3-luc (Aldosterone)
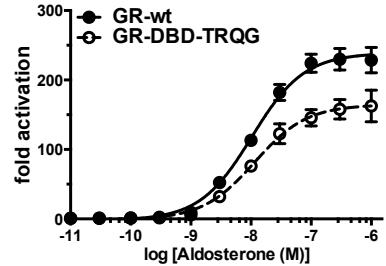

**Fig. 2. Concentration-dependent transcriptional activation by corticosteroids of wild-type and mutated elephant shark GR.** Plasmids for wild-type elephant shark GR and mutated elephant shark GR (with a TRQG insert) were expressed in HEK293 cells with either an MMTV-luciferase promoter (9,36,37) or a TAT3-luciferase promoter (9,38). Cells were treated with increasing concentrations of either aldosterone, cortisol, corticosterone, 11-deoxycortisol, 11-deoxycorticosterone or vehicle alone (DMSO). Results are expressed as means ± SEM, n=3. Y-axis indicates fold-activation compared to the activity of vector with vehicle (DMSO) alone as 1. Figure 2A-E. Elephant shark GRs with MMTV-luc. Figure 2F-J. Elephant shark GRs with TAT3-luc.

Elephant shark MR

**A: MMTV-luc (Cortisol)**
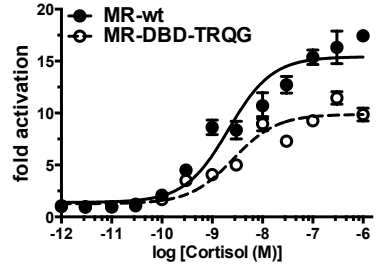

**F: TAT3-luc (Cortisol)**
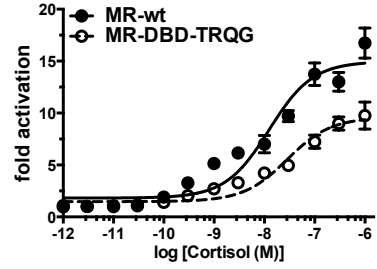

**B: MMTV-luc (11-deoxycorticosterone)**
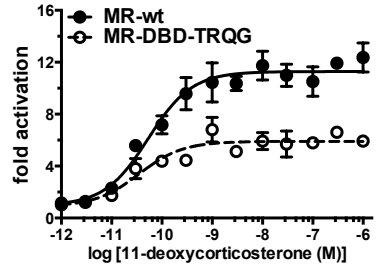

**G: TAT3-luc (11-deoxycorticosterone)**
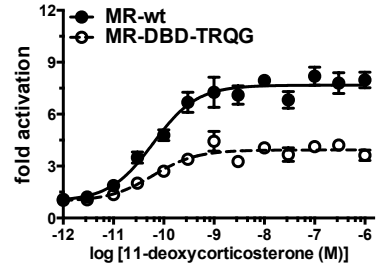

**C: MMTV-luc (Corticosterone)**
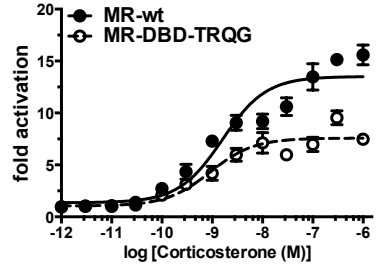

**H: TAT3-luc (Corticosterone)**
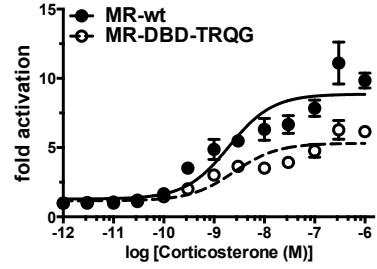

**D: MMTV-luc (11-deoxycortisol)**
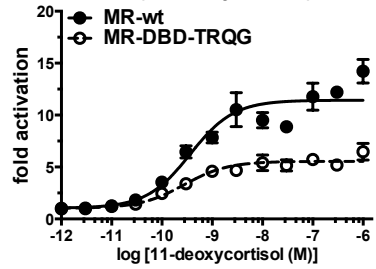

**I: TAT3-luc (11-deoxycortisol)**
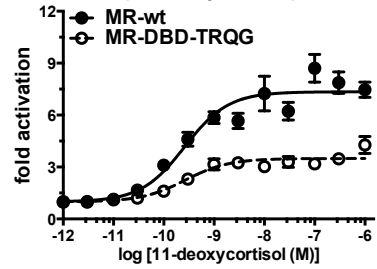

**E: MMTV-luc (Aldosterone)**
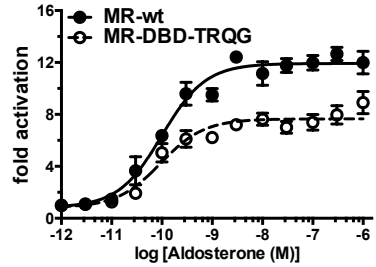

**J: TAT3-luc (Aldosterone)**
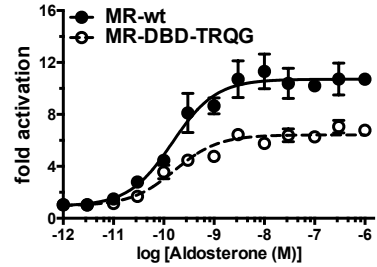

**Fig. 3. Concentration-dependent transcriptional activation by corticosteroids of wild-type and mutated elephant shark MR.** Plasmids for wild-type elephant shark MR and mutated elephant shark MR (with a TRQG insert) were expressed in HEK293 cells with either an MMTV-luciferase promoter or a TAT3-luciferase promoter. Cells were treated with increasing concentrations of either aldosterone, cortisol, corticosterone, 11-deoxycortisol, 11-deoxycorticosterone or vehicle alone (DMSO). Results are expressed as means ± SEM, n=3. Y-axis indicates fold-activation compared to the activity of vector with vehicle (DMSO) alone as 1. Figure 3A-E. Elephant shark MRs with MMTV-luc. Figure 3F-J. Elephant shark MRs with TAT3-luc.

**Table 1. Corticosteroid Activation of Wild-Type and Mutated Elephant Shark GR and MR in HEK293 Cells with either an MMTV Promoter or a TAT3 Promoter.**

| MMTV-luc | | Cortisol | DOC | Corticosterone | 11-deoxycortisol | Aldo |
|---|---|---|---|---|---|---|
| Elephant Shark GR -wt | EC50 (nM) | 34.9 | 12.5 | 5.1 | Curve did not saturate | 11.5 |
| | Fold-Activation (± SEM) | 580 (± 22.6) | 283 (± 12.2) | 602 (± 54.1) | Curve did not saturate) | 578 (± 36.3) |
| Elephant Shark GR -DBD-TRGQ | EC50 (nM) | 32.0 | 9.7 | 3.4 | Curve did not saturate) | 8.35 |
| | Fold-Activation (± SEM) | 381 (± 17.4) | 150 (± 2.1) | 285 (± 49.6) | Curve did not saturate) | 385 (± 12.5) |
| Elephant Shark MR -wt | EC50 (nM) | 2.1 | 0.053 | 1.6 | 0.36 | 0.1 |
| | Fold-Activation (± SEM) | 17.4 (± 0.5) | 12.4 (± 0.9) | 15.6 (± 0.9) | 14.2 (± 1.1) | 12.0 (± 0.9) |
| Elephant Shark MR -DBD-TRGQ | EC50 (nM) | 2.6 | 0.035 | 0.9 | 0.27 | 0.094 |
| | Fold-Activation (± SEM) | 9.9 (± 0.6) | 5.9 (± 0.2) | 7.5 (± 0.4) | 6.5 (± 0.8) | 8.9 (± 0.9) |
| **TAT3-luc** | | **Cortisol** | **DOC** | **Corticosterone** | **11-deoxycortisol** | **Aldo** |
| Elephant Shark GR -wt | EC50 (nM) | 27.1 | 12.9 | 5.0 | Curve does not saturate | 10.3 |
| | Fold-Activation (± SEM) | 157 (± 9.7) | 43.4 (± 1.4) | 189 (± 28.9) | Curve does not saturate | 229 (± 18.4) |
| Elephant Shark GR -DBD-TRGQ | EC50 (nM) | 45.7 | 16.5 | 5.0 | Curve does not saturate | 11.5 |
| | Fold-Activation (± SEM) | 141 (± 15.0) | 27.1 (± 2.1) | 109 (± 9.7) | Curve does not saturate | 163 (± 22.7) |
| Elephant Shark MR -wt | EC50 (nM) | 12.4 | 0.062 | 2.0 | 0.26 | 0.15 |
| | Fold-Activation (± SEM) | 16.7 (± 1.5) | 8.0 (± 0.5) | 9.8 (± 0.5) | 7.5 (± 0.4) | 10.7 (± 0.3) |
| Elephant Shark MR -DBD-TRGQ | EC50 (nM) | 28.5 | 0.06 | 2.5 | 0.27 | 0.17 |
| | Fold-Activation (± SEM) | 9.8 (± 1.3) | 3.6 (± 0.3) | 6.2 (± 0.1) | 4.3 (± 0.5) | 6.8 (± 0.1) |

wt = wild type sequence, DBD-TRGQ = TRGQ motif insertion in DBD

DOC = 11-deoxycorticosterone, Aldo = aldosterone

Curves for 11-deoxycortisol activation of elephant shark GR did not saturate at $10^{-6}$ M.

## Discussion

The sequencing of lamprey germline genome (35) led to the discovery of two CRs, which differ only in a small four residue deletion in the DBD of CR2 (8). In lamprey this deletion in CR2 reduces transcriptional activation by corticosteroids. Both elephant shark MR and GR lack this four amino acid insert in their DBDs, which raised the question of whether adding this four-residue insert to elephant shark MR and GR would reduce corticosteroid-mediated transcription, as seen in lamprey CR1 (8). As reported here (Figures 2 and 3, Table 1), we find that like lamprey CR1, addition of four residues -TRQG- to wild-type elephant shark MR and GR leads to a reduction in transcriptional activation by a panel of corticosteroids consisting of cortisol, corticosterone, aldosterone, 11-deoxycorticosterone and 11-deoxycortisol. This effect of the TRQG insert in elephant shark MR and GR is found for HEK293 cells transfected with either the MMTV or TAT3 promoters (Table 1). The loss of the TRQG insert in elephant shark GR and MR appears to have provided increased transcriptional activity for their GR and MR.

Meijsing et al (41) found that the function of the DBD in human GR was more than just docking of the GR to DNA. They showed that the human GR DBD also had an effect on transcription by human GR. Our finding that the DBD in elephant shark MR and GR and in sea lamprey CR has a role in regulating corticosteroid-mediated transcription indicates that a regulatory function of the DBD appeared early in the evolution of corticosteroid receptors.

## Methods

### Construction of plasmid vectors

Full-length glucocorticoid receptor (GR) and mineralocorticoid receptor (MR) sequences of elephant shark, *Callorhinchus milii*, were registered in Genbank (accession number: XP_042195980 for GR and XP_007902220 for MR). The insertion of 4-amino acids (Thr-Arg-Gln-Gly) into the DBD of elephant shark MR and GR was performed using KOD-Plus-mutagenesis kit (TOYOBO). The nucleic acid sequences of all constructs were verified by sequencing.

### Chemical reagents

Cortisol, corticosterone, 11-deoxycorticosterone, 11-deoxycortisol, and aldosterone were purchased from Sigma-Aldrich. For reporter gene assays, all hormones were dissolved in dimethyl-sulfoxide (DMSO); the final DMSO concentration in the culture medium did not exceed 0.1%.

**Transactivation assays and statistical analyses**

Transfection and reporter assays were carried out in HEK293 cells, as described previously (8). The cells were transfected with 100 ng of receptor gene, reporter gene containing the *Photinus pyralis* luciferase gene and pRL-tk, as an internal control to normalize for variation in transfection efficiency; pRL-tk contains the *Renilla reniformis* luciferase gene with the herpes simplex virus thymidine kinase promoter. Each assay had a similar number of cells, and assays were done with the same batch of cells in each experiment. All experiments were performed in triplicate. Promoter activity was calculated as firefly (*P. pyralis*)-lucifease activity/sea pansy (*R. reniformis*)-lucifease activity. The values shown are mean ± SEM from three separate experiments, and dose-response data, which were used to calculate the half maximal response (EC50) for each steroid, were analyzed using GraphPad Prism.

**Author Contributions**

**Conceptualization:** Yoshinao Katsu, Michael E. Baker.

**Data curation:** Yoshinao Katsu, Jiawen Zhang

**Formal analysis:** Yoshinao Katsu, Michael E. Baker.

**Investigation:** Yoshinao Katsu, Jiawen Zhang

**Methodology:** Yoshinao Katsu.

**Supervision:** Yoshinao Katsu, Michael E. Baker.

**Writing – original draft:** Yoshinao Katsu, Michael E. Baker.

**Writing – review & editing:** Yoshinao Katsu, Michael E. Baker.

**Funding:** This work was supported by Grants-in-Aid for Scientific Research from the Ministry of Education, Culture, Sports, Science and Technology of Japan (19K067309 to Y.K.), and the Takeda Science Foundation (to Y.K.).

## References

1. Osório J, Rétaux S. The lamprey in evolutionary studies. Dev Genes Evol. 2008 May;218(5):221-35. doi: 10.1007/s00427-008-0208-1. Epub 2008 Feb 15. PMID: 18274775.

2. Shimeld SM, Donoghue PC. Evolutionary crossroads in developmental biology: cyclostomes (lamprey and hagfish). Development. 2012 Jun;139(12):2091-9. doi: 10.1242/dev.074716. PMID: 22619386.

3. Smith JJ, Kuraku S, Holt C, Sauka-Spengler T, Jiang N, Campbell MS, Yandell MD, Manousaki T, Meyer A, Bloom OE, Morgan JR, Buxbaum JD, Sachidanandam R, Sims C, Garruss AS, Cook M, Krumlauf R, Wiedemann LM, Sower SA, Decatur WA, Hall JA, Amemiya CT, Saha NR, Buckley KM, Rast JP, Das S, Hirano M, McCurley N, Guo P, Rohner N, Tabin CJ, Piccinelli P, Elgar G, Ruffier M, Aken BL, Searle SM, Muffato M, Pignatelli M, Herrero J, Jones M, Brown CT, Chung-Davidson YW, Nanlohy KG, Libants SV, Yeh CY, McCauley DW, Langeland JA, Pancer Z, Fritzsch B, de Jong PJ, Zhu B, Fulton LL, Theising B, Flicek P, Bronner ME, Warren WC, Clifton SW, Wilson RK, Li W. Sequencing of the sea lamprey (Petromyzon marinus) genome provides insights into vertebrate evolution. Nat Genet. 2013 Apr;45(4):415-21, 421e1-2. doi: 10.1038/ng.2568. Epub 2013 Feb 24. PMID: 23435085; PMCID: PMC3709584.

4. Nakatani Y, Shingate P, Ravi V, Pillai NE, Prasad A, McLysaght A, Venkatesh B. Reconstruction of proto-vertebrate, proto-cyclostome and proto-gnathostome genomes provides new insights into early vertebrate evolution. Nat Commun. 2021 Jul 23;12(1):4489. doi: 10.1038/s41467-021-24573-z. Erratum in: Nat Commun. 2021 Jul 29;12(1):4704. PMID: 34301952; PMCID: PMC8302630.

5. Thornton JW. Evolution of vertebrate steroid receptors from an ancestral estrogen receptor by ligand exploitation and serial genome expansions. Proc Natl Acad Sci U S A. 2001 May 8;98(10):5671-6. doi: 10.1073/pnas.091553298. Epub 2001 May 1. PMID: 11331759; PMCID: PMC33271.

6. Carroll SM, Bridgham JT, Thornton JW. Evolution of hormone signaling in elasmobranchs by exploitation of promiscuous receptors. Mol Biol Evol. 2008;25(12):2643-2652. doi:10.1093/molbev/msn204.

7. Baker ME, Katsu Y. 30 YEARS OF THE MINERALOCORTICOID RECEPTOR: Evolution of the mineralocorticoid receptor: sequence, structure and function. J Endocrinol. 2017;234(1):T1-T16. doi:10.1530/JOE-16-0661.

8. Katsu Y, Lin X, Ji R, Chen Z, Kamisaka Y, Bamba K, Baker ME. N-terminal domain influences steroid activation of the Atlantic sea lamprey corticoid receptor. J Steroid Biochem Mol Biol. 2023 Jan 13;228:106249. doi: 10.1016/j.jsbmb.2023.106249. Epub ahead of print. PMID: 36646152.

9. Evans RM. The steroid and thyroid hormone receptor superfamily. Science. 1988;240(4854):889-895. doi:10.1126/science.3283939.

10. Bridgham JT, Eick GN, Larroux C, et al. Protein evolution by molecular tinkering: diversification of the nuclear receptor superfamily from a ligand-dependent ancestor. PLoS Biol. 2010;8(10):e1000497. Published 2010 Oct 5. doi:10.1371/journal.pbio.1000497.

11. Beato M, Klug J. Steroid hormone receptors: an update. Hum Reprod Update. 2000 May-Jun;6(3):225-36. doi: 10.1093/humupd/6.3.225. PMID: 10874567.

12. Whitfield GK, Jurutka PW, Haussler CA, Haussler MR. Steroid hormone receptors: evolution, ligands, and molecular basis of biologic function. J Cell Biochem. 1999;Suppl 32-33:110-22. doi: 10.1002/(sici)1097-4644(1999)75:32+<110::aid-jcb14>3.0.co;2-t. PMID: 10629110.

13. Baker ME. Steroid receptors and vertebrate evolution. Mol Cell Endocrinol. 2019;496:110526. doi:10.1016/j.mce.2019.110526.

14. Katsu Y, Shariful IMD, Lin X, Takagi W, Urushitani H, Kohno S, Hyodo S, Baker ME. N-terminal Domain Regulates Steroid Activation of Elephant Shark Glucocorticoid and Mineralocorticoid Receptors. J Steroid Biochem Mol Biol. 2021 Feb 27:105845. doi: 10.1016/j.jsbmb.2021.105845. Epub ahead of print. PMID: 33652098.

15. Fonseca E, Machado AM, Vilas-Arrondo N, Gomes-Dos-Santos A, Veríssimo A, Esteves P, Almeida T, Themudo G, Ruivo R, Pérez M, da Fonseca R, Santos MM, Froufe E, Román-Marcote E, Venkatesh B, Castro LFC. Cartilaginous fishes offer unique insights into the evolution of the nuclear receptor gene repertoire in gnathostomes. Gen Comp Endocrinol. 2020 Sep 1;295:113527. doi: 10.1016/j.ygcen.2020.113527. Epub 2020 Jun 8. PMID: 32526329.

16. Shibata S. 30 YEARS OF THE MINERALOCORTICOID RECEPTOR: Mineralocorticoid receptor and NaCl transport mechanisms in the renal distal nephron. J Endocrinol. 2017;234(1):T35-T47. doi:10.1530/JOE-16-0669.

17. Lifton RP, Gharavi AG, Geller DS. Molecular mechanisms of human hypertension. Cell. 2001;104(4):545-556. doi:10.1016/s0092-8674(01)00241-0.

18. Lombes M, Kenouch S, Souque A, Farman N, Rafestin-Oblin ME. The Mineralocorticoid Receptor Discriminates Aldosterone from Glucocorticoids Independently of the 11 Beta-Hydroxysteroid Dehydrogenase. Endocrinology. 1994;135(3):834-840. Doi:10.1210/Endo.135.3.8070376.

19. Pascual-Le Tallec L, Lombès M. The mineralocorticoid receptor: a journey exploring its diversity and specificity of action. Mol Endocrinol. 2005 Sep;19(9):2211-21. doi: 10.1210/me.2005-0089. Epub 2005 Mar 31. PMID: 15802372.

20. Rossier BC, Baker ME, Studer RA. Epithelial sodium transport and its control by aldosterone: the story of our internal environment revisited. Physiol Rev. 2015;95(1):297-340. doi:10.1152/physrev.00011.2014.

21. Hanukoglu I, Hanukoglu A. Epithelial sodium channel (ENaC) family: Phylogeny, structure-function, tissue distribution, and associated inherited diseases. Gene. 2016;579(2):95-132. doi:10.1016/j.gene.2015.12.061.

22. Jaisser F, Farman N. Emerging Roles of the Mineralocorticoid Receptor in Pathology: Toward New Paradigms in Clinical Pharmacology. Pharmacol Rev. 2016;68(1):49-75. doi:10.1124/pr.115.011106.

23. Hawkins UA, Gomez-Sanchez EP, Gomez-Sanchez CM, Gomez-Sanchez CE. The ubiquitous mineralocorticoid receptor: clinical implications. Curr Hypertens Rep. 2012;14(6):573-580. doi:10.1007/s11906-012-0297-0.

24. de Kloet ER, Joëls M. Brain mineralocorticoid receptor function in control of salt balance and stress-adaptation. Physiol Behav. 2017 Sep 1;178:13-20. doi: 10.1016/j.physbeh.2016.12.045. Epub 2017 Jan 13. PMID: 28089704.

25. Gomez-Sanchez EP. Brain mineralocorticoid receptors in cognition and cardiovascular homeostasis. Steroids. 2014 Dec;91:20-31. doi: 10.1016/j.steroids.2014.08.014. PMID: 25173821; PMCID: PMC4302001.

26. Gomez-Sanchez CE, Gomez-Sanchez EP. The Mineralocorticoid Receptor and the Heart. Endocrinology. 2021 Nov 1;162(11):bqab131. doi: 10.1210/endocr/bqab131. PMID: 34175946.

27. de Kloet ER. From receptor balance to rational glucocorticoid therapy. Endocrinology. 2014;155(8):2754-2769. doi:10.1210/en.2014-1048.

28. Chrousos GP. Stress and sex versus immunity and inflammation. Sci Signal. 2010 Oct 12;3(143):pe36. doi: 10.1126/scisignal.3143pe36. PMID: 20940425.

29. Cain DW, Cidlowski JA. Immune regulation by glucocorticoids. Nat Rev Immunol. 2017;17(4):233-247. doi:10.1038/nri.2017.1.

30. Weikum ER, Knuesel MT, Ortlund EA, Yamamoto KR. Glucocorticoid receptor control of transcription: precision and plasticity via allostery. Nat Rev Mol Cell Biol. 2017;18(3):159-174. doi:10.1038/nrm.2016.152.

31. Gross KL, Lu NZ, Cidlowski JA. Molecular mechanisms regulating glucocorticoid sensitivity and resistance. Mol Cell Endocrinol. 2009;300(1-2):7-16. doi:10.1016/j.mce.2008.10.001.

32. Close DA, Yun SS, McCormick SD, Wildbill AJ, Li W. 11-deoxycortisol is a corticosteroid hormone in the lamprey. Proc Natl Acad Sci U S A. 2010 Aug 3;107(31):13942-7. doi: 10.1073/pnas.0914026107. Epub 2010 Jul 19. PMID: 20643930; PMCID: PMC2922276.

33. Smith JJ, Antonacci F, Eichler EE, Amemiya CT. Programmed loss of millions of base pairs from a vertebrate genome. Proc Natl Acad Sci U S A. 2009 Jul 7;106(27):11212-7. doi: 10.1073/pnas.0902358106. Epub 2009 Jun 26. PMID: 19561299; PMCID: PMC2708698.

34. Smith JJ, Saha NR, Amemiya CT. Genome biology of the cyclostomes and insights into the evolutionary biology of vertebrate genomes. Integr Comp Biol. 2010 Jul;50(1):130-7. doi: 10.1093/icb/icq023. Epub 2010 Apr 19. PMID: 21558194; PMCID: PMC3140258.

35. Smith JJ, Timoshevskaya N, Ye C, Holt C, Keinath MC, Parker HJ, Cook ME, Hess JE, Narum SR, Lamanna F, Kaessmann H, Timoshevskiy VA, Waterbury CKM, Saraceno C, Wiedemann LM, Robb SMC, Baker C, Eichler EE, Hockman D, Sauka-Spengler T, Yandell M, Krumlauf R, Elgar G, Amemiya CT. The sea lamprey germline genome provides insights into programmed genome rearrangement and vertebrate evolution. Nat Genet. 2018 Feb;50(2):270-277. doi: 10.1038/s41588-017-0036-1. Epub 2018 Jan 22. Erratum in: Nat Genet. 2018 Apr 19;: Erratum in: Nat Genet. 2018 Nov;50(11):1617. PMID: 29358652; PMCID: PMC5805609.

36. Beato M, Arnemann J, Chalepakis G, Slater E, Willmann T. Gene regulation by steroid hormones. J Steroid Biochem. 1987;27(1-3):9-14. doi: 10.1016/0022-4731(87)90288-3. PMID: 2826895.

37. Cato AC, Skroch P, Weinmann J, Butkeraitis P, Ponta H. DNA sequences outside the receptor-binding sites differently modulate the responsiveness of the mouse mammary tumour virus promoter to various steroid hormones. EMBO J. 1988 May;7(5):1403-10. PMID: 2842149; PMCID: PMC458390.

38. Iñiguez-Lluhí JA, Pearce D. A common motif within the negative regulatory regions of multiple factors inhibits their transcriptional synergy. Mol Cell Biol. 2000 Aug;20(16):6040-50. doi: 10.1128/mcb.20.16.6040-6050.2000. PMID: 10913186; PMCID: PMC86080.

39. Rupprecht R, Arriza JL, Spengler D, et al. Transactivation and synergistic properties of the mineralocorticoid receptor: relationship to the glucocorticoid receptor. Mol Endocrinol. 1993;7(4):597-603. doi:10.1210/mend.7.4.8388999.

40. Kiilerich P, Triqueneaux G, Christensen NM, et al. Interaction between the trout mineralocorticoid and glucocorticoid receptors in vitro. J Mol Endocrinol. 2015;55(1):55-68. doi:10.1530/JME-15-0002.

41. Meijsing SH, Pufall MA, So AY, Bates DL, Chen L, Yamamoto KR. DNA binding site sequence directs glucocorticoid receptor structure and activity. Science. 2009 Apr 17;324(5925):407-10. doi: 10.1126/science.1164265. PMID: 19372434; PMCID: PMC2777810.